\documentclass[aps,twocolumn,groupedaddress]{revtex4}
\usepackage{amssymb,amsmath}
\usepackage{graphicx,array}
\usepackage{dcolumn}
\usepackage{bm}

\begin{document}
\title{Oxygen vacancies in SrTiO$_{3}$ thin films at finite temperatures: A first-principles study}

\author{Zizhen Zhou}
\affiliation{School of Materials Science and Engineering, UNSW Australia, Sydney NSW 2052, Australia}

\author{Dewei Chu}
\affiliation{School of Materials Science and Engineering, UNSW Australia, Sydney NSW 2052, Australia}

\author{Claudio Cazorla}
\thanks{Corresponding Author}
\affiliation{Departament de F\'isica, Universitat Polit\`ecnica de Catalunya, Campus Nord B4-B5, 
E-08034 Barcelona, Spain}

\begin{abstract}
Epitaxially grown SrTiO$_{3}$ (STO) thin films are material enablers for a number of critical energy-conversion 
and information-storage technologies like electrochemical electrode coatings, solid oxide fuel cells and random 
access memories. Oxygen vacancies (${\rm V_{O}}$), on the other hand, are key defects to understand and tailor 
many of the unique functionalities realized in oxide perovskite thin films. Here, we present a comprehensive and 
technically sound \emph{ab initio} description of ${\rm V_{O}}$ in epitaxially strained (001) STO thin films.
The novelty of our first-principles study lies in the incorporation of lattice thermal excitations on the 
formation energy and diffusion properties of ${\rm V_{O}}$ over wide epitaxial strain conditions ($-4 \le \eta 
\le +4$\%). We found that thermal lattice excitations are necessary to obtain a satisfactory agreement between
first-principles calculations and the available experimental data on the formation energy of ${\rm V_{O}}$ for
STO thin films. Furthermore, it is shown that thermal lattice excitations noticeably affect the energy barriers 
for oxygen ion diffusion, which strongly depend on $\eta$ and are significantly reduced (increased) under 
tensile (compressive) strain, also in consistent agreement with the experimental observations. The present work 
demonstrates that for a realistic theoretical description of oxygen vacancies in oxide perovskite thin films 
is necessary to consider lattice thermal excitations, thus going beyond standard zero-temperature \emph{ab 
initio} approaches.
\end{abstract}

\maketitle

\begin{figure}[t]
\centerline{
\includegraphics[width=1.00\linewidth]{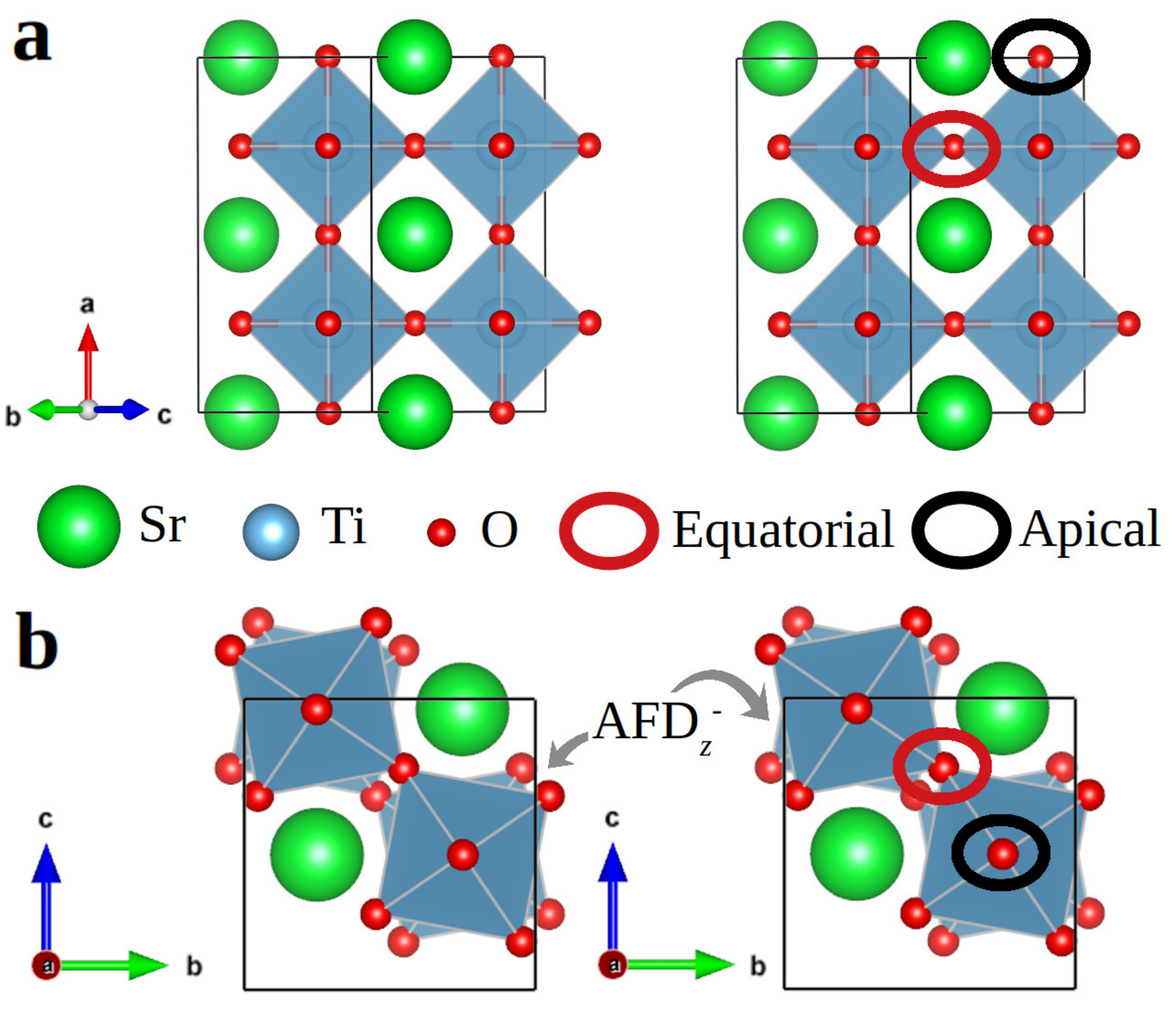}}
\caption{Representation of the $20$-atoms simulation cell employed for the simulation of stoichiometric 
	 and epitaixally strained (001) SrTiO$_{3}$ thin films. Different views of the simulation cell 
	 are represented in {\bf a} and {\bf b} along with the typical antiphase out-of-plane oxygen 
	 octahedral antiferrodistortive distortions (AFD$_{z}^{-}$) found in (001) SrTiO$_{3}$ thin films. 
	 The positions of generic equatorial and apical oxygen ions are also indicated in the figure.}
\label{fig1}
\end{figure}

\section{Introduction}
\label{sec:intro}

Crystalline defects, namely, deviations from the ideal and translationally invariant arrangement of atoms in 
crystals, are ubiquitous in real solids. Fortunately, the presence of crystalline defects is desirable for 
enhancing the functionality of many materials \cite{mofarah19a,mofarah19b,kong19,cazorla19}. A quintessential 
example of a functional type of crystalline defect are oxygen vacancies (${\rm V_{O}}$). Oxygen vacancies, for 
instance, can drastically boost the catalytic activity of transition metal oxide (TMO) surfaces by providing 
abundant reactive sites as well as highly mobile charges \cite{cazorla19,ganduglia-pirovano07,liu09}. The 
magnetic properties of TMO also can be altered substantially by changing their oxygen content since the 
exchange interactions between transition metal ions typically are sustained by O atoms \cite{hu18,menendez20,
lopez-bezanilla15}. Moreover, the presence of ${\rm V_{O}}$ enables ionic conductivity in perovskite-based 
solid solutions that are employed in critical electrochemical applications like solid oxide fuel and 
electrolysis cells \cite{huang06,zhou18}. Consequently, the functionality of TMO materials can be tailored 
and finely tuned through their stoichiometry. 
  
Another functionality design strategy that has proved very successful for TMO materials is strain engineering 
\cite{you19,liang19,schlom07}. Strain engineering consists in growing epitaxial thin films on top of substrates 
that present a lattice parameter mismatch, $\eta$. Either compressive or tensile biaxial stress can thus be 
introduced in the thin film upon the condition of coherent elastic coupling with the substrate. The ferroelectric 
\cite{schlom07,angsten17,cazorla15,oswaldo17,cazorla14}, magnetic \cite{menendez20,escorihuela12}, optical 
\cite{shi19,liu20} and catalytic \cite{you19,liu20} properties of TMO thin films can be drastically changed 
by strain engineering due to the existing strong couplings between their structural and electronic degrees of 
freedom. 

Recently, it has been realized that strain engineering also can be used to tune the formation and diffusion 
of oxygen vacancies in oxide thin films, and that such a combined $\eta$--${\rm V_{O}}$ physico-chemical approach 
represents a very promising technique for engineering new materials \cite{kerklotz17,jeen13,petrie16,hu16}. An 
illustrative example of the rich interplay between epitaxial strain and oxygen vacancies, which in turn may 
enormously influence the prevalent orbital and structural order parameters, is provided by the archetypal oxide 
perovskite SrTiO$_{3}$ (STO). 

Bulk STO is a quantum paraelectric crystal with an ideal cubic perovskite structure and high dielectric 
constant that strongly depends on temperature \cite{muller79}. Bulk STO is broadly used as a substrate 
in which to grow epitaxial perovskite thin films of high quality and as a key component of oxide heterostructures 
that exhibit fundamentally intriguing physical behaviour like LaTiO$_{3}$/STO bilayers (e.g., formation 
of a 2D electron gas at the interface) \cite{ohtomo04} and PbTiO$_{3}$/STO superlattices (e.g., emergence 
of polar vortices) \cite{yadav16}. At room temperature, STO epitaxial thin films may display in-plane 
ferroelectricity upon biaxial tensile stress \cite{haeni04} and out-of-plane ferroelectricity upon biaxial 
compressive stress \cite{yamada15}. Within a certain range of biaxial compressive stress, antiferrodistortive 
(AFD) oxygen octahedra rotations are observed to coexist with out-of-plane electric polarization \cite{yamada15}, 
thus pointing to the presence of unusual cooperative couplings between such normally opposing order parameters 
\cite{gu18}. 

The interplay between oxygen vacancies and biaxial stress in STO thin films is very rich and appears 
to be further enhanced by the coexistence of AFD distortions and ferroelectricity \cite{choi13,choi15}. 
Experimentally, it has been shown that both compressive and tensile biaxial strains significantly 
decrease the formation enthalpy of oxygen vacancies in STO thin films (e.g., by $\sim 10$\% for 
$|\eta| \sim 1$\%) \cite{iglesias17}. This behaviour is different from zero-temperature first-principles 
(also referred to as \emph{ab initio}) results obtained for other prototype oxide perovskites like 
BaTiO$_{3}$ and PbTiO$_{3}$, which indicate that compressive (tensile) biaxial stress typically 
depletes (promotes) the formation of ${\rm V_{O}}$ \cite{yang13a,yang13b}. Meanwhile, atomic force 
microscopy experiments have shown that tensile biaxial strain produces a substantial increase in the 
mobility of oxygen vacancies whereas small compressive biaxial strain produces an incipient ${\rm V_{O}}$ 
diffusion depletion \cite{iglesias18}. These latter experimental observations appear to be in partial 
disagreement with previous zero-temperature first-principles studies in which it has been concluded 
that both tensile and compressive biaxial strains tend to promote the migration of oxygen vacancies 
in STO thin films \cite{alhamadany13a,alhamadany13b}.     

First-principles calculations have been used to rationalize the atomistic mechanisms of strain-mediated 
formation of oxygen vacancies for a number of oxide perovskite thin films like BaTiO$_{3}$ \cite{yang13a}, 
PbTiO$_{3}$ \cite{yang13b}, CaMnO$_{3}$ \cite{aschauer13}, and SrCoO$_{3}$ \cite{hu16,cazorla17a}. As
previously mentioned, zero-temperature \emph{ab initio} calculations, which by definition neglect thermal
excitations, mostly agree in that the formation of ${\rm V_{O}}$ is strongly enhanced (reduced) by tensile 
(compressive) epitaxial strain. This behaviour has been explained in terms of an effective decrease in the 
electrostatic repulsive interactions between electronically reduced TM ions, which follows from an increase 
in the average distance between them \cite{aschauer13}. However, the nonmonotonic peak-like $\eta$-dependence 
of the ${\rm V_{O}}$ formation enthalpy that has been measured for biaxially strained STO thin films \cite{iglesias17} 
cannot be satisfactorily explained in terms of such electrostatic arguments (as otherwise the formation energy 
of oxygen vacancies should increase, rather than decrease, under compressive $\eta$ conditions). Likewise, the 
partial disagreements between theory and experiments on the diffusion properties of oxygen vacancies \cite{iglesias18,
alhamadany13a,alhamadany13b} appear to suggest that some key elements might be missing in the \emph{ab initio} 
calculations \cite{cazorla17a}.    

Here, we present a comprehensive first-principles study on the formation energy and migration of oxygen vacancies 
in epitaxially strained (001) STO thin films that explicitly incorporates lattice thermal effects. This improvement 
is achieved by means of quasi-harmonic free energy approaches and \emph{ab initio} molecular dynamics (AIMD) 
simulations (Sec.\ref{sec:methods}). In particular, we compare the formation energies and energy barriers for 
oxygen diffusion estimated both at zero temperature and $T \neq 0$ conditions, and discuss their agreement with 
the available experimental data. It is found that thermal lattice excitations are necessary to qualitatively 
reproduce the measured dependence of the ${\rm V_{O}}$ formation energy on biaxial stress. Thermal lattice 
excitations are also found to enhance ${\rm V_{O}}$ migration by reducing the involved energy barriers in about 
$40$\%. In agreement with the experiments, the diffusion coefficient of oxygen ions is found to substantially 
increase under tensile biaxial stress and to decrease under compressive biaxial stress. The present works 
evidences the need to use finite-temperature first-principles methods to rationalize the experimental findings 
on off-stoichiometric oxide perovskite thin films and to guide the engineering of new functional materials 
based on combined physico-chemical approaches.

\section{Computational methods}
\label{sec:methods}

\subsection{General technical details}
\label{subsec:comput}
We used the generalised gradient approximation to density functional theory (DFT) due to Perdew, Burke,
and Ernzerhof (GGA-PBE) \cite{pbe} as is implemented in the VASP software \cite{vasp}. A ``Hubbard-$U$'' 
scheme \cite{hubbard} was employed for a better treatment of Ti $3d$ electrons ($U_{\rm eff} = 2.0$~eV). 
We used the ``projector augmented wave'' method \cite{paw} to represent the ionic cores and considered 
the following electronic states as valence: Sr $4s$, $4p$ and $5s$; Ti $3p$, $4s$ and $3d$; O $2s$ and 
$2p$. Wave functions were represented in a plane-wave basis truncated at $650$~eV. For simulation of the 
stoichiometric systems, we employed a $20$-atoms simulation cell that allows to reproduce the usual 
ferroelectric and O$_{6}$ antiferrodistortive (AFD) distortions in perovskite oxides \cite{menendez20,cazorla15} 
(Fig.1). Off-stoichiometric systems containing oxygen vacancies, ${\rm V_{O}}$, were generated by removing 
oxygen atoms from either equatorial (Eq) or apical (Ap) positions (Fig.1). Simulation cells of different 
sizes were considered in order to quantify the effects of oxygen vacancy concentration on the obtained 
formation energy results. In particular, the following compositions were investigated: Sr$_{4}$Ti$_{4}$O$_{11}$ (or, 
equivalently, SrTiO$_{2.75}$), Sr$_{8}$Ti$_{8}$O$_{23}$ (SrTiO$_{2.88}$), and Sr$_{16}$Ti$_{16}$O$_{47}$ 
(SrTiO$_{2.94}$). For integrations within the Brillouin zone (BZ), we used a $\Gamma$-centered ${\bf k}$-point 
grid of $6 \times 8 \times 8$ for the $20$-atoms simulation cell and scaled it conveniently to maintain an 
equivalent ${\bf k}$-point density for the rest of cases. All oxygen vacancies were assumed to be neutrally
charged and non-magnetic (${\rm V_{O}}$) since this configuration has been shown to render the lowest energy 
for bulk off-stoichiometric SrTiO$_{3}$ in previous DFT studies \cite{lopez-bezanilla15,zhang16}.  

The geometry relaxations of epitaxially strained (001) SrTiO$_{3}$ and SrTiO$_{3-\delta}$ were carried out 
by using a conjugated gradient algorithm that allows to change the simulation-cell volume and atomic positions 
while constraining the length and orientation of the two in-plane lattice vectors (that is, $|a|=|b|$ and 
$\gamma = 90^{\circ}$). Periodic boundary conditions were applied along the three lattice-vector directions, 
thus the influence of surface and interface effects were systematically neglected in our simulations. This 
type of calculations are known as ``strained-bulk'' geometry relaxations and typically are considered to be 
a good approximation for thin films presenting thicknesses of at least few nanometers (that is, for which 
surface and interface effects can be safely disregarded) \cite{menendez20,cazorla15}. The simulated systems 
were assumed to be elastically coupled to a substrate thus the existence of possible stress relaxation 
mechanisms in the thin films were also neglected. The geometry relaxations were stopped when the forces on the 
ions were smaller than $0.01$~eV/\AA. By using these parameters we obtained zero-temperature energies that were 
converged to within $0.5$~meV per formula unit.

The electric polarization of stoichiometric and off-stoichiometric (001) SrTiO$_{3}$ thin films were estimated 
with the Born effective charges method \cite{menendez20,cazorla15}. In this approach, the electric polarization
is calculated via the formula:
\begin{equation} 
P_{\alpha} = \frac{1}{\Omega} \sum_{\kappa\beta}  Z_{\kappa\beta \alpha}^{*} u_{\kappa\beta}~, 
\label{eq:polarization}
\end{equation}
where $\Omega$ is the volume of the cell, $\kappa$ runs over all the atoms, $\alpha,\beta = x, y, z$ represent 
the Cartesian directions, $\bf{u}_{\kappa}$ is the displacement vector of the $\kappa$-th atom as referred to 
a non-polar reference phase, and $\boldsymbol{Z}^{*}_{\kappa}$ the Born effective charge tensor calculated for 
a non-polar reference state. It is worth noting that the presence of oxygen vacancies typically induced a notable 
reduction in the energy band gap of off-stoichiometric systems, which in some cases led to the appearence of 
metallic states. Consequently, estimation of the electric polarization with the more accomplished and accurate 
Berry phase formalism was not possible for all the analyzed compositions and thus we opted for systematically 
using the approximate Born effective charges method \cite{menendez20,menendez20b}.

\subsection{Phonon calculations}
\label{subsec:phonons}
To estimate phonon frequencies we employed the ``small-displacement'' approach \cite{phon}, in which the 
force-constant matrix of the crystal is calculated in real space by considering the proportionality between 
the atomic displacements and forces when the former are sufficiently small (in the present study this 
condition was satisfied for atomic displacements of $0.02$~\AA). Large supercells containing $160$ atoms were 
employed to guarantee that the elements of the force-constant matrix presented practically negligible values 
at the largest atomic separations. We used a dense ${\bf k}$-point grid of $3 \times 3 \times 3$ for the 
calculation of the atomic forces with VASP. The computation of the nonlocal parts of the pseudopotential 
contributions were performed in reciprocal space in order to maximise the numerical accuracy. Once a 
force-constant matrix was determined, we Fourier transformed it to obtain the phonon frequencies for any 
arbitrary ${\bf k}$-point in the first BZ. This latter step was performed with the PHON code \cite{phon}, in which 
the translational invariance of the system is exploited to ensure that the three acoustic branches are exactly 
zero at the $\Gamma$ point. Central differences for the atomic forces, that is, both positive and negative 
atomic displacements, were considered. A complete phonon calculation involved the evaluation of atomic forces 
for $120$ ($114$) different stoichiometric (off-stoichiometric) configurations with the technical parameters 
just described. In order to accurately compute $F^{\rm qh}_{\rm vac}$ (see below), we employed a dense 
${\bf k}$-point grid of $16 \times 16 \times 16$ for BZ integration. With these settings we found that the 
calculated quasi-harmonic free energies were accurate to within $5$~meV per formula unit.

\subsection{Free energy calculations}
\label{subsec:quasiharm}
We computed the quasi-harmonic Gibbs free energy associated with the formation of neutral oxygen vacancies, 
$G^{\rm qh}_{\rm vac}$, as a function of epitaxial strain, $\eta \equiv \left(a - a_{0}\right) / a_{0}$ (where 
$a_{0}$ represents the equilibrium in-plane lattice parameter calculated for the stoichiometric system), and 
temperature, $T$, by following the approach introduced in previous works \cite{hu16,cazorla17a}. Next, we 
briefly summarize the key aspects and technical details of the employed quasi-harmonic Gibbs free energy 
calculation method.

The formation Gibbs free energy of non-magnetic and neutrally charged ${\rm V_{O}}$ can be expressed as 
\cite{hu16,cazorla17a}: 
\begin{equation}
G^{\rm qh}_{\rm vac}(\eta, T) = E_{\rm vac}(\eta) + F^{\rm qh}_{\rm vac}(\eta, T) + \mu_{\rm O}(T)~, 
\label{eq1}
\end{equation}
where subscript ``vac'' indicates the quantity difference between the off-stoichiometric and stoichiometric 
systems (e.g., $E_{\rm vac} \equiv E_{\rm SrTiO_{3-\delta}} - E_{\rm SrTiO_{3}}$), $E_{\rm vac}$ accounts for 
the static contributions to the free energy (i.e., calculated at $T = 0$ conditions by considering the atoms 
fixed at their equilibrium lattice positions \cite{menendez20}), $F^{\rm qh}_{\rm vac}$ for the vibrational 
contributions to the free energy, and $\mu_{\rm O}$ is the chemical potential of free oxygen atoms. The 
vibrational free energy of stoichiometric and off-stoichiometric systems were estimated with the quasi-harmonic 
formula \cite{cazorla13,cazorla09,cazorla15b,baroni09,cazorla17}:  
\begin{equation}
 F^{\rm qh}(\eta, T) = \frac{1}{N_{q}}~k_{\rm B} T \sum_{{\bf
    q}s}\ln\left[ 2\sinh \left( \frac{\hbar\omega_{{\bf
        q}s}(\eta)}{2k_{\rm B}T} \right) \right]~,  
\label{eq2}
\end{equation}
where $N_{q}$ is the total number of wave vectors used for integration within the Brillouin zone and the 
dependence of the phonon frequencies, $\omega_{\boldsymbol{q}s}$, on epitaxial strain is explicitly noted. 

It is well known that first-principles estimation of $\mu_{\rm O}$ with DFT$+U$ methods is challenging and 
may lead to large errors \cite{jones89,wang06}. Such inherent limitations make the prediction of ${\rm V_{O}}$ 
formation energies by exclusively using DFT approaches difficult and probably also imprecise. Notwithstanding, 
since (i)~the oxygen chemical potential does not depend on epitaxial strain (i.e., in practice the value of 
$\mu_{\rm O}$ is determined by the experimental conditions) and (ii)~our main aim is to unravel the impact 
of lattice excitations on the formation energy and diffusion of ${\rm V_{O}}$ as a function of $\eta$, we can 
safely base our analysis on the results obtained for the thermodynamically shifted Gibbs free energy:
\begin{equation}
G^{* \rm qh}_{\rm vac}(\eta, T) = G^{\rm qh}_{\rm vac}(\eta, T) - \mu_{\rm O}(T)~.
\label{eq3}
\end{equation}
In other words, rather than adopting experimental values for $\mu_{\rm O}$ and/or applying empirical 
corrections to the calculated vacancy formation energies \cite{aschauer13,wang06}, we select arbitrary
values for the oxygen-gas chemical potential without any loss of generality.

\subsection{Nudged elastic band calculations}
\label{subsec:neb}
\emph{Ab initio} nudged-elastic band (NEB) calculations \cite{henkelman00} were performed to estimate the 
activation energy for ${\rm V_{O}}$ diffusion in expitaxially strained (001) SrTiO$_{3}$ at zero temperature. 
Our NEB calculations were performed for reasonably large $2 \times 2 \times 2$ or $3 \times 3 \times 3$ 
supercells containing several tens of atoms \cite{zhang16}. We used ${\bf q}$-point grids of $8 \times 8 
\times 8$ or $6 \times 6 \times 6$ and an energy plane-wave cut-off of $650$~eV. Six intermediate images 
were used to determine the most likely ${\rm V_{O}}$ diffusion paths in the absence of thermal excitations. 
The geometry optimizations were halted when the total forces on the atoms were smaller than 
$0.01$~eV$\cdot$\AA$^{-1}$. The NEB calculations were performed for five expitaxial-strain equidistant points 
in the interval $-4 \le \eta \le 4$\%.

\subsection{\emph{Ab initio} molecular dynamics simulations}
\label{subsec:aimd}
First-principles molecular dynamics (AIMD) simulations based on DFT were performed in the canonical 
$(N, V, T)$ ensemble. The selected volumes and geometries were those determined at zero-temperature 
conditions, hence we neglected thermal expansion effects. The concentration of oxygen vacancies in 
the off-stoichiometric systems was also considered to be independent of $T$ and equal to $\approx 
1.6$\%. The temperature in the AIMD simulations was kept fluctuating around a set-point value by 
using Nose-Hoover thermostats. Large simulation boxes containing $317$ atoms (Sr$_{64}$Ti$_{64}$O$_{189}$) 
were employed in all the AIMD simulations and periodic boundary conditions were applied along the three 
Cartesian directions. Newton's equation of motion were integrated by using the customary Verlet's 
algorithm and a time-step length of $\delta t = 10^{-3}$~ps. $\Gamma$-point sampling for integration 
within the first Brillouin zone was employed in all the AIMD simulations. The calculations comprised 
total simulation times of $t_{total} \sim 10$~ps. We performed three AIMD simulations at $T = 1000$, 
$1500$, and $2000$~K for off-stoichiometric STO thin films considering epitaxial strains of $-3.6$, 
$0$ and $+3.6$\%. 

The mean square displacement (MSD) of oxygen ions was estimated with the formula \cite{sagotra19}:
\begin{eqnarray}
{\rm MSD}(\tau) & = & \frac{1}{N_{ion} \left( N_{step} - n_{\tau} \right)} \times \\ \nonumber
                &   & \sum_{i=1}^{N_{ion}} \sum_{j=1}^{N_{step} - n_{\tau}} | {\bf r}_{i} (t_{j} + \tau) - {\bf r}_{i} (t_{j}) |^{2}~,
\label{msd}
\end{eqnarray}
where ${\bf r}_{i}(t_{j})$ is the position of a migrating ion $i$ at time $t_{j}$ ($= j \cdot \delta t$), 
$\tau$ represents a lag time, $n_{\tau} = \tau / \delta t$, $N_{ion}$ is the total number of mobile ions, 
and $N_{step}$ the total number of time steps. The maximum $n_{\tau}$ was chosen equal to $N_{step}/3$ 
(i.e., equivalent to $\sim 3$--$4$~ps), hence we could accumulate enough statistics to reduce significantly 
the ${\rm MSD}(\tau)$ fluctuations at the largest $\tau$ (see the error bars in the ${\rm MSD}$ plots 
presented in the following sections). Oxygen diffusion coefficients were subsequently obtained with the 
Einstein relation:
\begin{equation}
D =  \lim_{\tau \to \infty} \frac{{\rm MSD}(\tau)}{6\tau}~.
\label{dcoeff}
\end{equation}
The $T$-dependence of the oxygen diffusion coefficient was assumed to follow the Arrhenius formula:
\begin{equation}
        D(T) = D_{0} \cdot \exp{\left[-\frac{E_{a}}{k_{B}T} \right]}~,
\label{arrhenius}
\end{equation}
where $D_{0}$ is known as the pre-exponential factor, $E_{a}$ is the activation energy for ionic diffusion,
and $k_{B}$ the Boltzmann constant.

\begin{figure*}[t]
\centerline{
\includegraphics[width=1.00\linewidth]{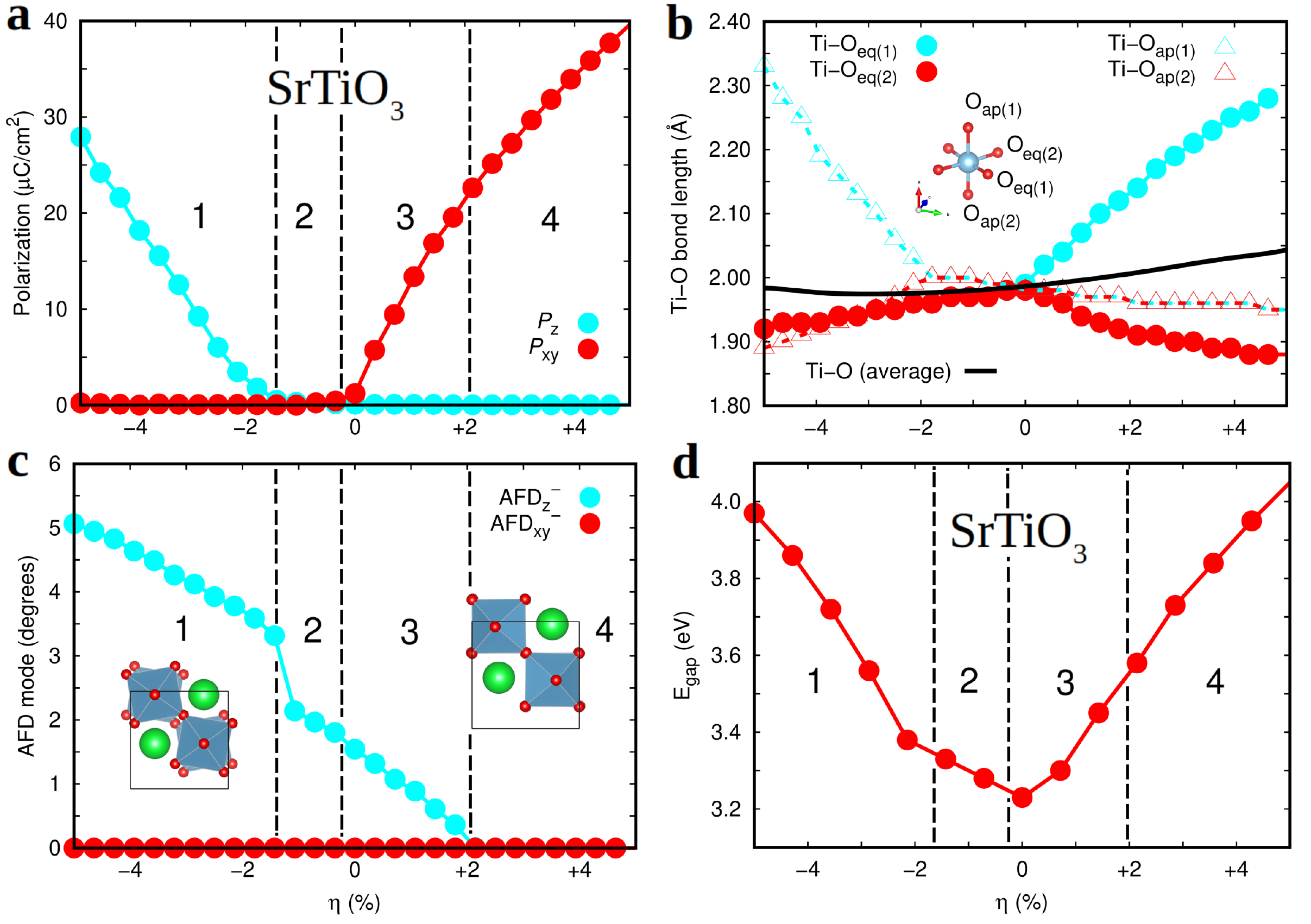}}
\caption{Physical properties of stoichiometric (001) SrTiO$_{3}$ thin films estimated with
	 first-principles methods. {\bf a} The electric polarization along the out-of-plane ($P_{z}$) 
	 and in-plane ($P_{xy}$) directions expressed as a function of expitaxial strain, $\eta$. 
	 {\bf b} Different Ti--O bond lengths expressed as a function of epitaxial strain. {\bf c} 
         Antiphase out-of-plane (AFD$_{z}^{-}$) and in-plane (AFD$_{xy}^{-}$) antiferrodistortive 
         O$_{6}$ rotations expressed as a function of epitaxial strain. {\bf d} Energy band gap of 
         (001) SrTiO$_{3}$ thin films estimated with the range-separated hybrid functional HSE06 
         \cite{hse06} on the geometries determined at the PBE+$U$ level. The vertical lines indicate 
         different phase stability regions, namely, 1.~$I4cm$, 2.~$I4/mcm$, 3.~$Ima2$ and 4.~$Amm2$.}
\label{fig2}
\end{figure*}

\section{Results and Discussion}
\label{sec:results}
We start by discussing the zero-temperature phase diagram of stoichiometric (001) STO thin films calculated 
with first-principles methods. The changes in the structural and electric polarization properties 
induced by the presence of equatorial (Eq) and apical (Ap) oxygen vacancies (${\rm V_{O}}$) are subsequently     
explained. The impact of thermal effects on the formation energy of ${\rm V_{O}}$ is analyzed for a wide
range of epitaxial strain ($\eta$) and temperature conditions. We also report and compare the energy barriers 
for ionic oxygen diffusion in (001) STO thin films estimated by neglecting and by taking into account $T$-induced 
lattice vibrations. Insightful connections between our theoretical results and experimental measurements are 
provided whenever the latter are available in the literature.

\subsection{Zero-temperature properties of stoichiometric (001) SrTiO$_{3}$ thin films}
\label{subsec:stoi}
Figure~\ref{fig2} shows the structural, electric polarization and energy band gap properties of stoichiometric 
(001) SrTiO$_{3}$ thin films estimated with first-principles methods (i.e., density functional theory --DFT--, 
Sec.~\ref{sec:methods}) at zero temperature. Different crystalline phases are stabilized as a result of varying
the epitaxial strain conditions, which are described in detail next. We note that several authors have previously
reported analogous DFT results to ours \cite{angsten17,antons05,lin06,lebedev16} and that the best agreement 
with the present calculations is obtained for work \cite{lebedev16}, in which antiferrodistortive oxygen 
octahedra rotations (AFD) were also explicitly modeled.   
  
In the epitaxial strain interval $\eta \lesssim -2$\%, we observe the stabilization of a tetragonal $I4cm$
phase that is characterized by a significant out-of-plane polarization ($P_{z}$, Fig.\ref{fig2}a) and antiphase
out-of-plane O$_{6}$ rotations (AFD$_{z}^{-}$, Fig.\ref{fig2}c). Coexistence of the order parameters $P_{z}$ 
and AFD$_{z}^{-}$ is quite unique as they normally tend to oppose each other \cite{gu18}, a polar-antiferrodistortive 
interplay that has been experimentally observed and characterized as a function temperature and $\eta$ \cite{yamada15}. 
Under tensile strain, half of the Ti--O bond lengths involving oxygen atoms in apical positions are significantly 
elongated as compared to those involving O ions in equatorial positions (Fig.\ref{fig2}b), a structural distortion 
that signals the presence of out-of-plane polarization \cite{cazorla15,cazorla14}.   

In the epitaxial strain interval $-2 \lesssim \eta \lesssim 0$\%, a tetragonal $I4/mcm$ phase appears
that presents null electric polarization (Fig.\ref{fig2}a) and moderate antiphase out-of-plane O$_{6}$ 
rotations (Fig.\ref{fig2}c). In this phase, the length of the Ti--O bonds are all pretty similar regardless 
of the positions occupied by the oxygen atoms (Fig.\ref{fig2}b). It is worth noting that when some tiny 
monoclinic lattice distortions in the generated equilibrium geometries (i.e., $\alpha \sim 0.1$ degrees) 
are not disregarded the identification of this phase is also compatible with a non-polar $C2/c$ phase that 
is similar to the one previously predicted for metallic LaNiO$_{3}$ thin films \cite{weber16}.    

In the epitaxial strain interval $0 \lesssim \eta \lesssim +2$\%, a noticeable in-plane electric polarization,
$P_{xy}$, appears in the system that coexists with small AFD$_{z}^{-}$ O$_{6}$ rotations (Fig.\ref{fig2}a,c).
The resulting crystal phase is orthorhombic and its symmetry can be ascribed to the polar space group $Ima2$.
Under tensile strain, half of the Ti--O bond lengths involving oxygen atoms in equatorial positions are 
significantly elongated as compared to those involving O ions in apical positions (Fig.\ref{fig2}b), a 
structural distortion that produces a significant in-plane polarization \cite{cazorla15,cazorla14}.
In the epitaxial strain interval $\eta \gtrsim +2$\%, the antiphase out-of-plane O$_{6}$ rotations
completely disappear and $P_{xy}$ grows steadily under increasing epitaxial strain. In this latter case, 
the optimized crystal structure is also orthorhombic and its symmetry can be identified with the space group 
$Amm2$.    

Figure~\ref{fig2}d shows the energy band gap of (001) SrTiO$_{3}$ thin films, $E_{\rm gap}$, estimated as a 
function of epitaxial strain with the range-separated hybrid functional HSE06 \cite{hse06}. The reason for 
including this information here will become clearer in the next subsection, where we explain the oxygen 
vacancy formation energy results obtained at zero temperature. It is worth noting that $E_{\rm gap}$
increases noticeably under either tensile or compressive biaxial strain as compared to the corresponding 
zero-strain value. For instance, at $\eta = 0$ the energy band gap amounts to $3.2$~eV whereas at 
$\eta = \pm 4$\% is approximately equal to $3.9$~eV. Such a $\eta$-induced $E_{\rm gap}$ trend is markedly 
different from the one predicted for binary oxides like CeO$_{2}$ and TiO$_{2}$ by using analogous 
first-principles methods \cite{liu20}, which displays a significant $E_{\rm gap}$ reduction under tensile 
biaxial strain. The reason for such a difference in $E_{\rm gap}$ behaviour is likely to be related to the 
larger changes in the dielectric susceptibility that can be induced by epitaxial strain in STO thin films 
as compared to binary oxides \cite{liu20,antons05}.    

\begin{figure*}[t]
\centerline{
\includegraphics[width=1.00\linewidth]{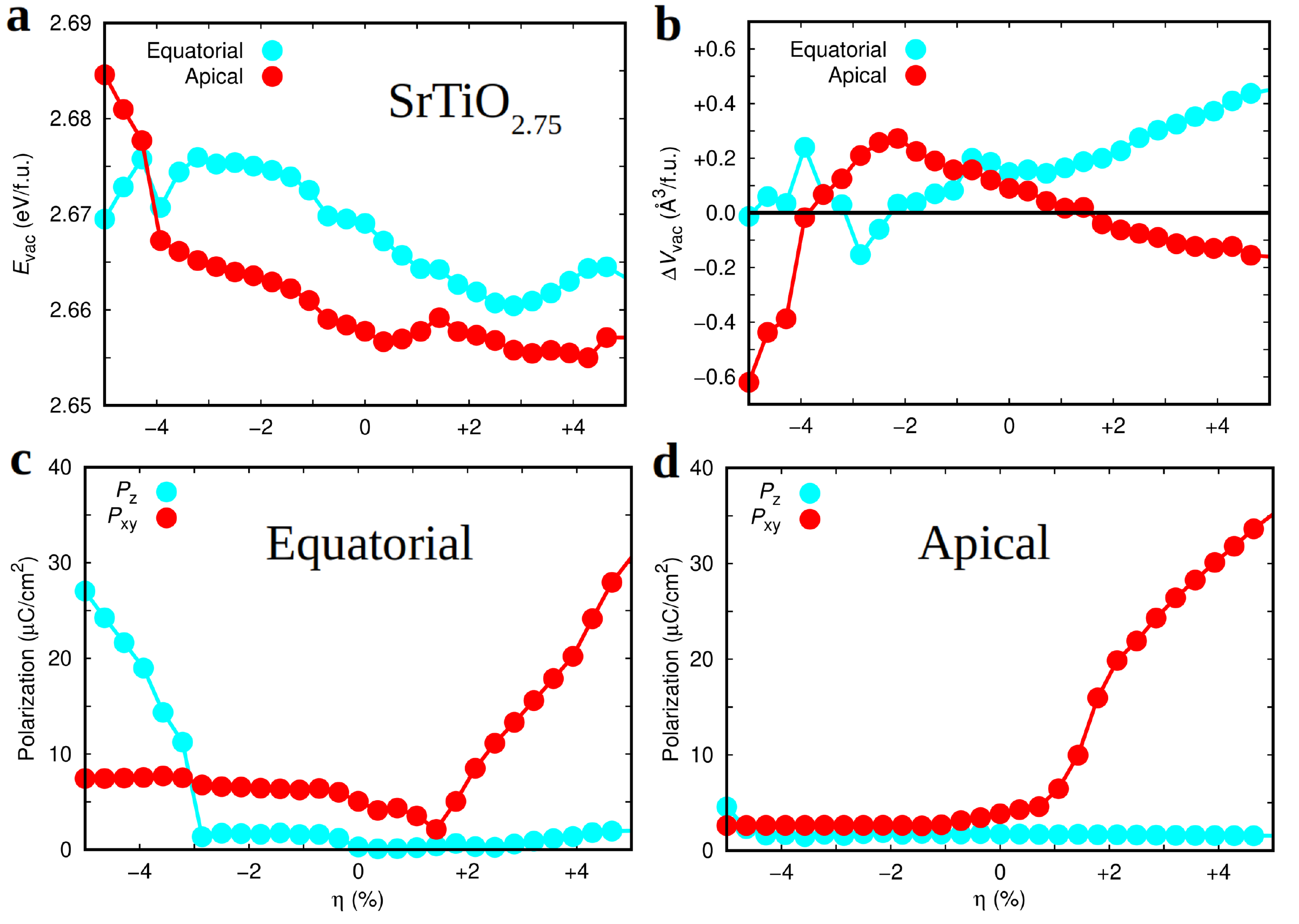}}
\caption{Zero-temperature properties of non-stoichiometric epitaxially strained (001) SrTiO$_{2.75}$ estimated 
	with first-principles methods based on DFT (Sec.~\ref{sec:methods}). {\bf a} Zero-temperature formation 
	energy of oxygen vacancies expressed as a function of oxygen position and epitaxial strain. {\bf b}
	Volume change per formula unit, $\Delta V_{\rm vac} \equiv V_{\rm SrTiO_{2.75}} - V_{\rm SrTiO_{3}}$,  
	induced by the creation of oxygen vacancies and expressed as a function of oxygen position and epitaxial 
	strain. The electric polarization along the out-of-plane ($P_{z}$) and in-plane ($P_{xy}$) directions 
	expressed as a function of expitaxial strain for (001) SrTiO$_{2.75}$ thin films containing {\bf c} 
	equatorial and {\bf d} apical oxygen vacancies.}
\label{fig3}
\end{figure*}

\subsection{Formation energy of oxygen vacancies at $T = 0$}
\label{subsec:zero-vac}
Figure~\ref{fig3} shows the formation energy of oxygen vacancies calculated for (001) STO thin films at
zero temperature, $E_{\rm vac}$. The concentration of ${\rm V_{O}}$ considered in this case renders the 
composition SrTiO$_{2.75}$ (analogous vacancy energy results obtained for smaller oxygen vacancy concentrations 
are explained below). A small decrease in $E_{\rm vac}$ is observed as the biaxial strain changes from 
compressive to tensile (i.e., of $> 1$\% when considering the two limiting cases $\eta = \pm 4$\%, 
Fig.\ref{fig3}a). For most $\eta$ cases, it seems more favourable to create apical ${\rm V_{O}}$ 
than equatorial, however the formation energy differences between the two cases are pretty small (i.e., 
$|\Delta E_{\rm vac}| \approx 0.01$~eV per formula unit, Fig.\ref{fig3}a).    

For negative $\eta$ values, the creation of oxygen vacancies in either equatorial (Fig.\ref{fig3}c) or
apical (Fig.\ref{fig3}d) positions has a dramatic effect on the electric polarization of the system.
In particular, the sizable out-of-plane polarization found in stoichiometric STO thin films (Fig.\ref{fig2}a) 
practically disappears when ${\rm V_{O}}$ are exclusively created in apical positions. Meanwhile, when
oxygen vacancies are generated solely in equatorial positions a non-negligible in-plane polarization 
of $\approx 7$~$\mu$C~cm$^{-2}$ appears for any value of compressive epitaxial strain. For positive $\eta$ 
values, on the other hand, the general behaviour of the electrical polarization is quite similar to that 
found for the analogous stoichiometric thin films, although the size of $P_{xy}$ appreciably decreases 
($\sim 10$\%).        

Figure~\ref{fig3}b shows the volume difference between SrTiO$_{2.75}$ and stoichiometric thin films,
$\Delta V_{\rm vac}$, expressed as a function of epitaxial strain. The creation of neutral oxygen vacancies 
in oxide perovskites typically induces an increase in volume, the so-called chemical expansion, due to the
electronic reduction of transition metal ions that are located close to ${\rm V_{O}}$'s \cite{cazorla17a,
marrocchelli15}. For present purposes, it is interesting to analyze the $\eta$--dependence of $\Delta 
V_{\rm vac}$ because this quantity has been found to be correlated with the contribution of lattice 
thermal excitations to the formation energy of ${\rm V_{O}}$ at finite temperatures \cite{cazorla17a}. As 
regards equatorial oxygen vacancies, $\Delta V_{\rm vac}$ turns out to be positive and moderately large (small) 
under tensile (compressive) epitaxial strain. By contrast, the creation of apical oxygen vacancies is 
accompanied by negative (positive) and large $\Delta V_{\rm vac}$ absolute values (small) at large compressive 
(tensile) epitaxial strain (Fig.\ref{fig3}b). In the next subsection, we will comment on possible correlations 
between these zero-temperature $\Delta V_{\rm vac}$ results and the lattice-related contributions to the 
formation energy of ${\rm V_{O}}$ at finite temperatures (i.e., the Gibbs free energy $G^{* \rm qh}_{\rm vac}$ 
shown in Eq.(\ref{eq3})). 
         
\begin{figure}[t]
\centerline{
\includegraphics[width=1.00\linewidth]{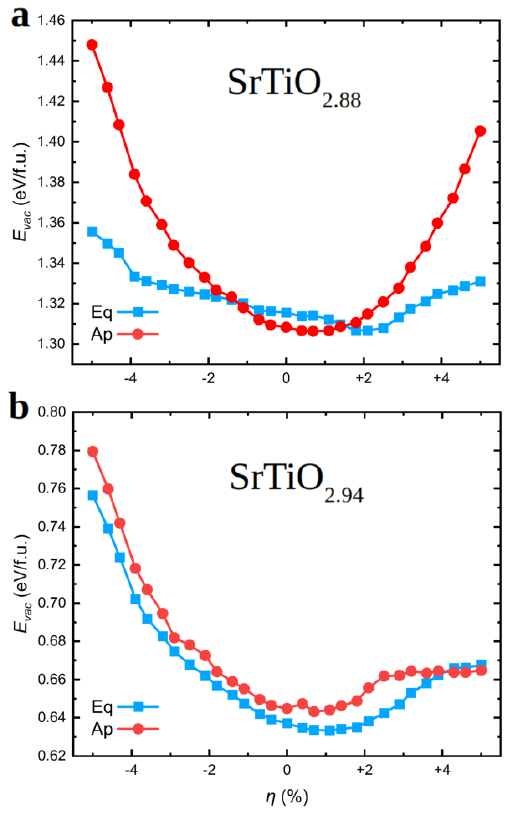}}
\caption{Zero-temperature formation energy of oxygen vacancies expressed as a function of oxygen position and 
	epitaxial strain for compositions {\bf a} SrTiO$_{2.88}$ and {\bf b} SrTiO$_{2.94}$. Labels ``Eq'' and 
	``Ap'' stand out for equatorial and apical ${\rm V_{O}}$, respectively.}
\label{fig4}
\end{figure}

The estimation of oxygen vacancy formation energies may depend strongly on the concentration of ${\rm V_{O}}$ 
considered in the simulations due to the presence of short- and long-ranged interactions acting between the 
defects \cite{mofarah19b,menendez20}. Figures~\ref{fig3}a and \ref{fig4} explicitly show this effect, as it 
is found that by decreasing the ${\rm V_{O}}$ concentration the computed zero-temperature formation energy 
dramatically decreases for any arbitrary value of $\eta$. For instance, the estimated $E_{\rm vac}$ for 
unstrained SrTiO$_{2.75}$ and SrTiO$_{2.94}$ amounts to $2.7$ and $0.6$~eV, respectively. This result suggests 
that short and middle-range interactions between oxygen vacancies are of repulsive type and thus the 
formation of ${\rm V_{O}}$ clusters in STO thin films in principle is not likely to occur at low and moderate 
temperatures. Moreover, the $E_{\rm vac}$ difference between equatorial and apical oxygen vacancies also 
depends critically on the concentration of defects. Specifically, according to our $E_{\rm vac}$ results obtained
for SrTiO$_{2.75}$ thin films in general it is more favourable to create apical ${\rm V_{O}}$ than equatorial 
(Fig.\ref{fig3}a) whereas for SrTiO$_{2.94}$ thin films the tendency is just the opposite (Fig.\ref{fig4}b).
The effect of epitaxial strain on $E_{\rm vac}$ also varies as the concentration of ${\rm V_{O}}$ changes. 
In particular, $E_{\rm vac}$ increases both under compressive and tensile strains for SrTiO$_{2.94}$ thin 
films whereas for SrTiO$_{2.75}$ it decreases under tensile strain. 

How do these zero-temperature ${\rm V_{O}}$ formation energy results compare with the available experimental 
data? In a recent paper, Rivadulla and collaborators have measured the enthalpy of oxygen vacancy formation 
for STO thin films as a function of epitaxial stress \cite{iglesias17}. The authors have found that under both 
compressive and tensile strains such energy noticeably decreases. For instance, in the experiments the 
${\rm V_{O}}$ formation enthalpy decreases by $\approx 20$\% ($\approx 40$\%) for a tensile (compressive) 
strain of $1$\% as compared to the unstrained case \cite{iglesias17}. Therefore, the agreement between our 
zero-temperature $E_{\rm vac}$ results expressed as a function of $\eta$ and ${\rm V_{O}}$ concentration 
(Figs.\ref{fig3}a,\ref{fig4}) and the experimental observations is far from satisfactory. In order to
fundamentally understand the origins of such large discrepancies, and based on the fact that oxygen vacancies
in oxide perovskites typically are created at high temperatures \cite{hu18,hu16,iglesias17}, we proceeded
to explicitly calculate ${\rm V_{O}}$ formation free energies at finite temperatures (rather than at 
non-realistic $T = 0$ conditions).        

Before explaining our ${\rm V_{O}}$ formation energy results obtained at $T \neq 0$ conditions, it is worth 
mentioning that in a recent work \cite{choi15} another first-principles study on the ${\rm V_{O}}$ formation 
energy of STO thin films has been reported. Zero-temperature $E_{\rm vac}$ results analogous to ours are 
presented in \cite{choi15}, however, the conclusions reported in that study are drastically different from 
the computational outcomes just described in this section. In particular, a systematic decrease in $E_{\rm vac}$
has been predicted for either tensile or compressive strains, which is the opposite behaviour than what 
we have found here for SrTiO$_{2.94}$ thin films, for instance. Moreover, an intriguing correlation between 
the $\eta$-induced behaviour of $E_{\rm vac}$ and the energy band gap of STO thin films ($E_{\rm gap}$) has 
been also suggested in work \cite{choi15}. Based on our results enclosed in Figs.\ref{fig2}d and \ref{fig4}b, 
such a correlation is partially corroborated \cite{liu20}. Nevertheless, in our calculations both quantities 
$E_{\rm vac}$ and $E_{\rm gap}$ increase, rather than decrease, under either tensile or compressive strains. 
We hypothesize that the likely reasons for such theoretical disagreements may be the neglection of 
characteristic STO structural motifs in work \cite{choi15}, like polar and antiferrodistortive oxygen 
octahedral distortions.               

\begin{figure}[t]
\centerline{
\includegraphics[width=1.00\linewidth]{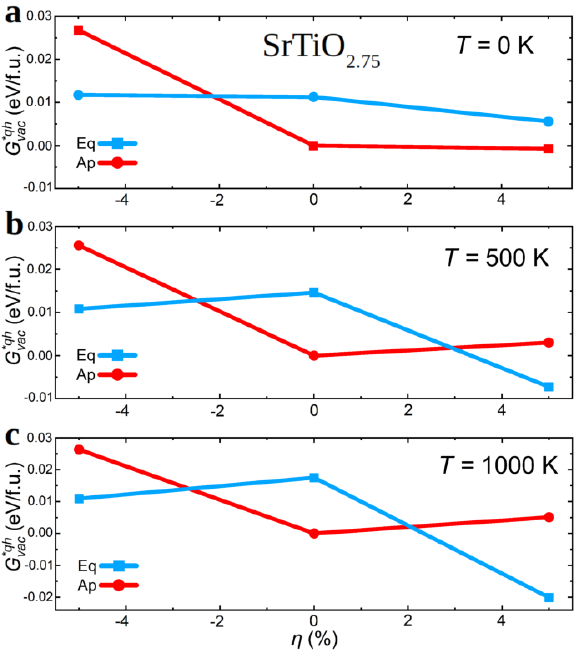}}
\caption{Thermodynamically $\mu_{\rm O}$-shifted Gibbs free energy [Eqs.(\ref{eq1})--(\ref{eq3})] (Sec.
         \ref{subsec:quasiharm}) for ${\rm V_{O}}$ formation expressed as a function of oxygen vacancy 
         position, epitaxial strain and temperature. {\bf a} $T = 0$~K, {\bf b} $T = 500$~K, and {\bf c} 
         $T = 1000$~K. The composition of the off-stoichiometric (001) thin films corresponds to 
         SrTiO$_{2.75}$ and labels ``Eq'' and ``Ap'' stand out for equatorial and apical oxygen vacancies, 
         respectively.}
\label{fig5}
\end{figure}

\subsection{Formation energy of oxygen vacancies at $T \neq 0$}
\label{subsec:gibbs}
We performed quasi-harmonic Gibbs free energy calculations to estimate the formation energy of oxygen 
vacancies at finite temperatures, $G^{* \rm qh}_{\rm vac}$ [Eq.(\ref{eq3})], for epitaxially constrained 
(001) SrTiO$_{2.75}$ thin films using the methods explained in Sec.\ref{subsec:quasiharm}. Unfortunately,
due to the huge computational effort associated with the calculation of phonon spectra of off-stoichiometric 
systems, we could not assess the dependence of $G^{* \rm qh}_{\rm vac}$ on the concentration of 
${\rm V_{O}}$. Figure~\ref{fig5} shows our $G^{* \rm qh}_{\rm vac}$ results expressed as as function of 
temperature and epitaxial strain. Since here we are primarily interested in analyzing the joint effects 
of epitaxial strain and lattice thermal excitations on the formation energy of oxygen vacancies, the chemical 
potential entering Eq.(\ref{eq3}) has been arbitrarily selected, without any loss of generality, to provide 
null $G^{* \rm qh}_{\rm vac}$ values for the minimum energy determined under $\eta = 0$ conditions at each 
temperature. 

We found that lattice thermal excitations hardly affect the $\mu_{\rm O}$-shifted formation energy of apical 
${\rm V_{O}}$'s, independently of the epitaxial strain. In particular, only a small $G^{* \rm qh}_{\rm vac}$ 
increase of few meV per formula unit is appreciated under tensile strain as compared to the values 
estimated at zero temperature (Figs.\ref{fig5}a,c). By contrast, the $\eta$-dependence of the $\mu_{\rm 
O}$-shifted formation energy of equatorial ${\rm V_{O}}$'s drastically changes as a result of considering 
$T$-induced lattice vibrations. For instance, at the highest analyzed temperature, $T =1000$~K, $G^{* \rm 
qh}_{\rm vac}$ decreases by as much as $\approx 50$\% for a biaxial strain of $-5$\% and by $\approx 
200$\% for $\eta = +5$\% (Fig.\ref{fig5}c). For an intermediate temperature of $500$~K, the observed tendency 
is analogous to the one just described although the $G^{* \rm qh}_{\rm vac}$ differences with respect to the 
unstrained case are slightly smaller (i.e., a reduction of $\approx 35$\% and $\approx 150$\% for $\eta = -5$ 
and $+5$\%, respectively --Fig.\ref{fig5}b--).      

The differences between the estimated $G^{* \rm qh}_{\rm vac}$ as a function of $T$ and $\eta$ for apical 
and equatorial ${\rm V_{O}}$ can be qualitatively understood in terms of the zero-temperature proxy 
$\Delta V_{\rm vac}$ introduced in Sec.\ref{subsec:zero-vac} (Fig.\ref{fig3}b). In a recent theoretical 
paper \cite{cazorla17a}, it has been proposed that for positive $\Delta V_{\rm vac}$ values, that is, 
$V_{\rm SrTiO_{3-\delta}} > V_{\rm SrTiO_{3}}$, lattice thermal excitations tend to facilitate the formation 
of oxygen vacancies. As it is observed in Fig.\ref{fig3}b, for equatorial vacancies $\Delta V_{\rm vac}$ is
positive and steadily increases under tensile biaxial strain; this outcome is agreeing with the large relative 
$G^{* \rm qh}_{\rm vac}$ decrease estimated for $\eta = +5$\% upon increasing temperature (Fig.\ref{fig5}).
Meanwhile, for apical vacancies $\Delta V_{\rm vac}$ is negative under both large tensile and compressive 
strains; this behaviour is consistent with the fact that under increasing temperature the corresponding  
relative $G^{* \rm qh}_{\rm vac}$ differences hardly change. Therefore, we corroborate the previously 
proposed qualitative correlation between the two quantities $\Delta V_{\rm vac}$ and $F^{\rm qh}_{\rm 
vac}$ (Sec.\ref{subsec:quasiharm}), which are computed at zero temperature and $T \neq 0$ conditions, 
respectively \cite{cazorla17a}. 

How do these finite-temperature ${\rm V_{O}}$ formation energy results compare with the experimental data
reported in work \cite{iglesias17}? The answer is that although the agreement between theory and observations
is not quantitative it can be regarded as qualitatively satisfactory. We recall that experimentally
it has been determined that under both compressive and tensile biaxial strains oxygen vacancies can be 
created more easily. This behaviour is analogous to what we have predicted for equatorial ${\rm V_{O}}$'s, 
which in oxide perovskites correspond to the most representative class of anion positions (i.e., equatorial 
O sites are 50\% more numerous than apical). Moreover, since the $G^{* \rm qh}_{\rm vac}$ values estimated 
for equatorial ${\rm V_{O}}$'s under both tensile and compressive biaxial strains are smaller than those 
estimated for apical vacancies (by $\approx 30$ and $20$ meV per formula unit, respectively), it is 
likely that to a certain extent vacancy ordering occurs in epitaxially strained STO thin films (as it 
has been experimentally shown for grain boundaries in bulk STO from scanning transmission electron microscopy 
measurements \cite{clie01}). On the down side, experiments indicate that it is more easy to create oxygen 
vacancies under compressive strain than under tensile strain \cite{iglesias17} while our calculations predict 
the opposite trend (Fig.\ref{fig5}). Nonetheless, based on our computational $E_{\rm vac}$ and $G^{* \rm 
qh}_{\rm vac}$ results, it can be concluded that in order to reproduce the experimentally observed 
$\eta$-induced enhancement of ${\rm V_{O}}$ formation with theoretical \emph{ab initio} methods it is 
necessary to explicitly consider vibrational lattice thermal excitations in the calculations.  

\begin{figure*}[t]
\centerline{
\includegraphics[width=1.00\linewidth]{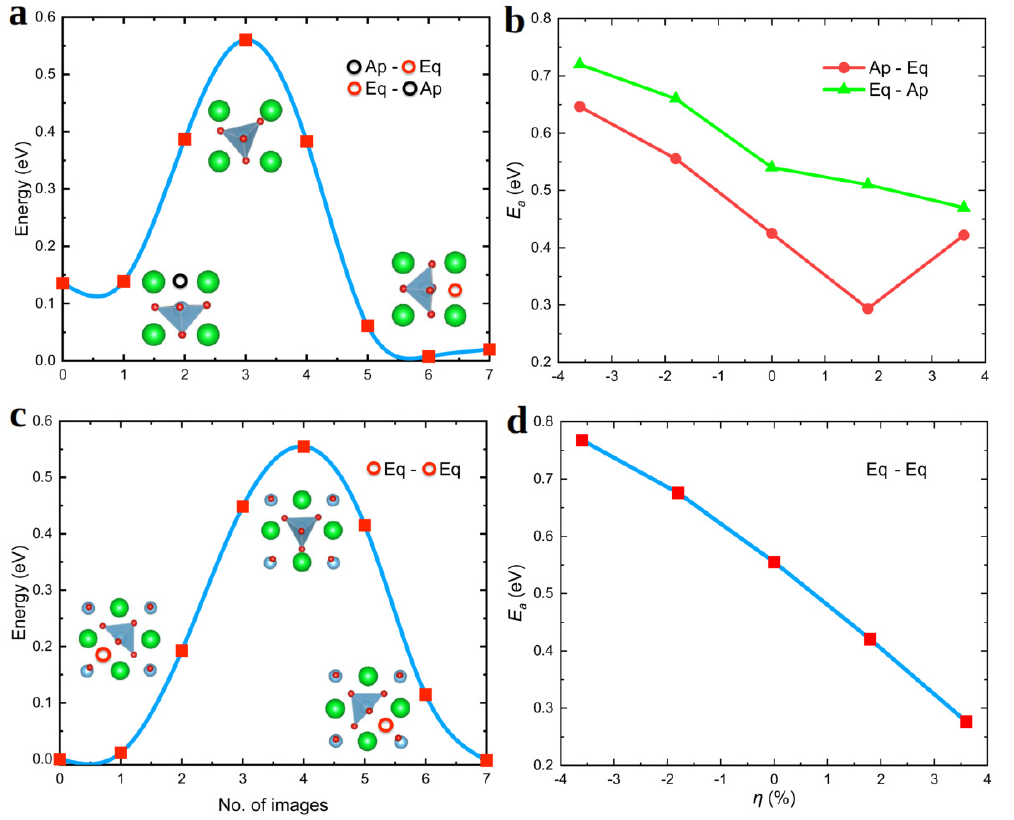}}
\caption{Energy barriers for ${\rm V_{O}}$ diffusion calculated with the NEB method (Sec.\ref{subsec:neb}) and by 
	neglecting thermal lattice fluctuations. Representation of the analyzed oxygen vacancy diffusion paths 
	are shown in {\bf a} and {\bf c} ($\eta = 0$ case). NEB energy barrier results expressed as a function 
	of epitaxial strain are represented in {\bf b} and {\bf d}. Labels ``Eq'' and ``Ap'' stand out for 
	equatorial and apical oxygen vacancies, respectively. The colouring code for atoms in {\bf a} and 
	{\bf c} coincides with that indicated in Fig.\ref{fig1}.}
\label{fig6}
\end{figure*}

\begin{figure*}[t]
\centerline{
\includegraphics[width=1.00\linewidth]{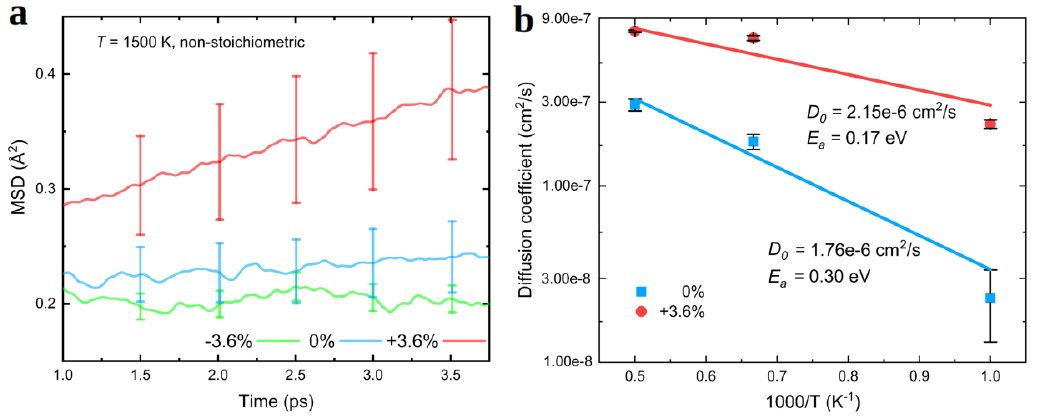}}
\caption{{\bf a} Mean square displacement (MSD) calculated for oxygen ions in off-stoichiometric (001) STO thin
        films with AIMD simulations performed at $T = 1500$~K and considering different epitaxial strain
        conditions, namely, $\eta = -3.6$, $0$ and $+3.6$\%. {\bf b} Oxygen diffusion coefficients estimated
        for off-stoichiometric (001) STO thin films with AIMD simulations considering different temperatures
        and epitaxial strain conditions. The resulting pre-exponential factors, $D_{0}$, and activation
        energies, $E_{a}$, for oxygen ionic diffusion are indicated in the plot (Sec.\ref{subsec:aimd}).}
\label{fig7}
\end{figure*}

\subsection{Zero-temperature activation energy for oxygen diffusion}
\label{subsec:neb-results}
The diffusion of ${\rm V_{O}}$'s in oxide perovskites is a key parameter for the design of ionic-based 
devices \cite{wan18}. In recent atomic force microscopy experiments performed by Iglesias \emph{et al.} 
\cite{iglesias18}, it has been shown that tensile biaxial strain produces a substantial increase in the 
diffusion of O ions in STO thin films. In particular, the room-temperature diffusion coefficient of oxygen 
atoms, $D_{\rm O}$, roughly increases by a factor of $4$ upon a tensile biaxial strain of $\approx +2$\% 
\cite{iglesias18}. For compressive tensile strains, on the other hand, the available experimental data 
is quite scarce. Nonetheless, measurements performed up to a $\eta$ of $\approx -1$\% appear to suggest 
an incipient reduction in $D_{\rm O}$ \cite{iglesias18}. On this regard, first-principles analysis of 
ionic transport properties may be very useful as calculations are free of the technical problems found 
in the experimental synthesis of epitaxially grown thin films and thus arbitrarily large tensile/compressive 
biaxial strains can be simulated. 

First-principles simulation of ionic diffusion processes, however, are neither exempt of some technical issues 
and shortcomings \cite{sagotra19}. For instance, due to the intense computational expense associated with $T 
\neq 0$ simulations, most first-principles studies usually neglect temperature effects. In particular, 
zero-temperature calculations of ion-migration energy barriers typically are performed with the nudged elastic 
band (NEB) method \cite{henkelman00} (Sec.\ref{subsec:neb-results}), in which (i)~the initial and final 
diffusion positions of the vacancy and interstitial ions need to be guessed in the form of high-symmetry 
configurations rendering metastable states, and (ii)~$T$-induced lattice excitations are totally 
neglected. Limitations of the NEB method for accurately determining ionic diffusion energy barriers and
paths are well known and documented for some prototype fast-ion conductor materials (e.g., see works 
\cite{sagotra19} and \cite{yang11}). 

Al-Hamadany \emph{et al.} have already studied the migration of oxygen vacancies in (001) STO thin films 
by means of NEB and DFT methods \cite{alhamadany13a,alhamadany13b}. For the case of tensile biaxial strains, 
Al-Hamadany \emph{et al.} have reported a systematic and significant reduction in the energy barrier for 
${\rm V_{O}}$ diffusion, $E_{a}$ (i.e., of up to $25$\% for large $\eta$'s of $+6$--$8$\% \cite{alhamadany13b}). 
This computational outcome is in good agreement with the experimental tendency found by Iglesias \emph{et 
al.} for $D_{\rm O}$ \cite{iglesias18}. The value of the reported NEB activation energy calculated at 
zero-strain conditions is approximately $0.8$~eV. For the case of compressive biaxial strains, Al-Hamadany 
\emph{et al.} have also reported a decrease in $E_{a}$ for high $|\eta|$'s of $> 4$\% \cite{alhamadany13a}
(i.e., of up to $50$\% for $\eta$'s of $-6$--$8$\%); in the $0 \le \eta \le 4$ interval, on the other hand, 
the energy barrier for ${\rm V_{O}}$ diffusion hardly changes or increases just moderately (depending on the 
considered initial and final oxygen vacancy positions).

Figure~\ref{fig6} shows our $E_{a}$ results obtained for (001) STO thin films by employing DFT NEB techniques
(Sec.\ref{subsec:neb-results}). Two possible ${\rm V_{O}}$ diffusion paths, namely, ``Ap-Eq'' (Fig.\ref{fig6}a) 
and ``Eq-Eq'' (Fig.\ref{fig6}c) where ``Ap'' and ``Eq'' stand for apical and equatorial O sites, have been 
considered in our simulations. In the former case, we obtain two different energy barriers, ``Ap-Eq'' and 
``Eq-Ap'', due to the energy asymmetry between the two involved oxygen positions (Figs.\ref{fig3} and \ref{fig4}).
In consistent agreement with the available experimental data and previous DFT studies, we find that under 
tensile biaxial strain the energy barrier for ${\rm V_{O}}$ diffusion is greatly reduced. For instance, at 
$\eta \approx +4$\% we obtain that $E_{a}$ decreases with respect to the value estimated at zero strain (i.e., 
$0.55$~eV) by $\approx 50$\% and $15$\% for ``Eq-Eq'' (Fig.\ref{fig6}d) and ``Eq-Ap'' (Fig.\ref{fig6}b), 
respectively. (The ${\rm V_{O}}$ diffusion energy barrier difference between cases ``Eq-Ap'' and ``Ap-Eq'' 
simply correspond to the zero-temperature ${\rm V_{O}}$ formation energy difference between cases ``Eq'' 
and ``Ap''.) It is worth noting that our estimated zero-strain $E_{a}$ value of $0.55$~eV is in very good 
agreement with the experimental ${\rm V_{O}}$ diffusion energy barrier measured for bulk STO, $E_{a}^{\rm expt} 
\approx 0.60$~eV \cite{souza12}.
 
Upon compressive biaxial strain, we find that $E_{a}$ increases significantly and practically linearly with 
$|\eta|$ (Fig.\ref{fig6}b,d). For instance, at $\eta \approx -4$\% we predict that $E_{a}$ increases with respect 
to the zero-strain value of $0.55$~eV by $\approx 45$\% and $32$\% for ``Eq-Eq'' (Fig.\ref{fig6}d) and ``Eq-Ap'' 
(Fig.\ref{fig6}b), respectively. These results appear to be in agreement with the scarce experimental data 
that is available for compressive biaxial strains \cite{iglesias18} but in clear disagreement with previous 
DFT results reported by Al-Hamadany \emph{et al.} \cite{alhamadany13a}. The reasons for the disagreements 
between our theoretical NEB $E_{a}$ estimations and others \cite{alhamadany13a} are not clear to us since the 
distinctive structural motifs of STO thin films (e.g., polar and antiferrodistortive oxygen octahedral 
distortions) were considered in all works. In order to fully test the reliability of our $E_{a}$ 
zero-temperature NEB results, we performed complementary \emph{ab initio} molecular dynamics (AIMD) simulations 
in which lattice thermal excitations are fully taken into account and no particular ${\rm V_{O}}$ diffusion 
path needs to be guessed \cite{sagotra19}.

\subsection{Oxygen ionic diffusion at finite temperature}
\label{subsec:diffusion}
Figure~\ref{fig7} encloses the MSD and $D_{\rm O}$ results obtained from our $T \neq 0$ AIMD simulations
for (001) STO thin films at $\eta = \pm 3.6$\% and zero strain (Sec.\ref{subsec:aimd}). For the $\eta = 
0$ case, we estimate large diffusion coefficients of $\sim 10^{-8}$--$10^{-7}$~cm$^{2}$s$^{-1}$ at 
temperatures higher than $1000$~K and a small ${\rm V_{O}}$ diffusion energy barrier of $0.30$~eV (Fig.\ref{fig7}b). 
The pre-exponential factor entering the corresponding $D_{\rm O}$ Arrhenius formula (Sec.\ref{subsec:aimd}) 
amounts to $1.8 \cdot 10^{-6}$~cm$^{2}$s$^{-1}$. The $E_{a}$ value estimated by fully considering lattice 
thermal excitations is approximately $50$\% smaller than the one calculated with the NEB method considering 
zero-temperature conditions. This computational outcome demonstrates the existence of an important interplay 
between lattice vibrations and ${\rm V_{O}}$ diffusion, which in the case of STO thin films enormously facilitates
ionic transport. It is also worth noting that the agreement between our zero-strain $E_{a}$ result obtained from 
AIMD simulations and the experimental diffusion energy barrier $E_{a}^{\rm expt} \approx 0.60$~eV \cite{souza12} 
has considerably worsened as compared to the corresponding NEB estimation. Possible causes explaining such 
an extended disagreement could be the neglection of other types of defects in our $T \neq 0$ calculations, 
like dislocations \cite{marrocchelli15b}, and the fact that the concentration of oxygen vacancies in our AIMD 
simulations ($\approx 1.6$\%) is probably larger than in the samples analyzed in the experiments. 

For a tensile strain of $+3.6$\%, we find that the diffusion of oxygen vacancies is considerably enhanced 
as compared to the $\eta = 0$ case. In particular, we estimate high-$T$ diffusion coefficients of $\sim 
10^{-7}$~cm$^{2}$s$^{-1}$ and a reduced ${\rm V_{O}}$ diffusion energy barrier of $0.17$~eV (Fig.\ref{fig7}b). 
The value of the pre-exponential factor entering the corresponding $D_{\rm O}$ Arrhenius formula 
(Sec.\ref{subsec:aimd}) is equal to $2.2 \cdot 10^{-6}$~cm$^{2}$s$^{-1}$. The $E_{a}$ decrease induced by 
$\eta = +3.6$\% is about $50$\% of the zero-strain value, which is very similar to the relative variation 
determined with NEB techniques for the same biaxial strain and ``Eq-Eq'' vacancy diffusion path 
(Sec.\ref{subsec:neb-results}). In this case, it is also concluded that the effects of lattice thermal 
excitations is to significantly enhance oxygen transport. 

As regards compressive biaxial strains, it is found that even at temperatures as high as $1500$ and $2000$~K 
the diffusion coefficient of oxygen atoms is nominally zero (Fig.\ref{fig7}a). This AIMD result is in qualitative 
agreement with the NEB calculations presented in the previous section, since in the latter case we found that $E_{a}$ increases almost linearly 
with $|\eta|$ (Sec.\ref{subsec:neb-results}). We note that if the energy barrier for ${\rm V_{O}}$ diffusion hardly changed 
under large compressive strains, then for $\eta = -3.6$\% we would have estimated similar MSD and $D_{\rm O}$ 
values to those obtained for the unstrained system, which is not the case.   

Overall, the AIMD simulation results presented in this section confirm the correctness (at the qualitative level) 
of our NEB results reported in Sec.\ref{subsec:neb-results}, and demonstrate that lattice thermal vibrations have a significant 
enhancing effect on ${\rm V_{O}}$ diffusion in (001) STO thin films. Interestingly, it is not always the case that 
lattice thermal excitations are found to promote ionic transport. For instance, in a recent systematic theoretical 
study on Li-based fast-ion conductors \cite{sagotra19} the opposite trend has been demonstrated, namely, the 
energy barriers for ionic transport estimated from AIMD simulations in general are higher than those obtained with 
NEB methods. It is likely that the degree of anharmonicity of the non-diffusing lattice in the considered
material, which determines the amplitude of the atomic fluctuations around the corresponding equilibrium 
positions, is directly related to the either enhancing or suppressing ionic diffusion effect mediated by the 
lattice excitations. Further quantitative investigations on this subject deserve future work.

\section{Conclusions}
\label{sec:conclusions}
We have presented a comprehensive \emph{ab initio} study on the formation energy and diffusion properties 
of oxygen vacancies in epitaxially strained (001) STO thin films, a class of functional materials with great 
fundamental and applied interests. The novelty of our work lies in the incorporation of lattice thermal 
excitations on the first-principles description of ${\rm V_{O}}$. It has been demonstrated that in order 
to achieve an improved agreement with the experimental observations it is necessary to explicitly consider 
temperature-induced lattice effects in the theoretical calculations. For instance, by performing quasi-harmonic 
Gibbs free energy calculations we have been able to qualitatively reproduce the nonmonotonic peak-like 
dependence of the ${\rm V_{O}}$ formation enthalpy measured in experiments. Also, by performing \emph{ab initio} 
molecular dynamics simulations we have been able to reproduce the qualitative $\eta$-driven ${\rm V_{O}}$
diffusion trends observed in biaxially strained (001) STO samples. Generalization of our main conclusions 
to other technologically relevant oxide perovskite materials is likely, although further experimental and 
computational works on the interplay between oxygen vacancies and epitaxial strain are necessary. We hope 
that the present study will stimulate research efforts in this direction.

\section*{Acknowledgments}
D.C. and C.C. acknowledge support from the Australian Research Council through funded projects LP190100113 
and LP190100829. C.C. acknowledges support from the Spanish Ministry of Science, Innovation and Universities 
under the ``Ram\'on y Cajal'' fellowship RYC2018-024947-I. Computational resources and technical assistance 
were provided by the Australian Government and the Government of Western Australia through Magnus under the 
National Computational Merit Allocation Scheme and The Pawsey Supercomputing Centre, and by the University
of Valencia through the Tirant III cluster and its technical service.

\end{document}